\documentstyle[emulateapj,psfig]{article}

\begin{document}
\def\ale{\mathrel{\hbox{\rlap{\hbox{\lower4pt\hbox{$\sim$}}}\hbox{$<$}}}}
\def\age{\mathrel{\hbox{\rlap{\hbox{\lower4pt\hbox{$\sim$}}}\hbox{$>$}}}}

\title{The Unusually Long Duration Gamma-ray Burst GRB~000911
\footnote{Partially based on observations obtained at the W.~M.~Keck
Observatory which is operated by the California Association for Research
in Astronomy, a scientific partnership among California
Institute of Technology, the University of California and
the National Aeronautics and Space Administration.}
}

\author{
P.~A.~Price\altaffilmark{1,2},
E.~Berger\altaffilmark{1},
S.~R.~Kulkarni\altaffilmark{1},
S.~G.~Djorgovski\altaffilmark{1},
D.~W.~Fox\altaffilmark{1},
A. Mahabal\altaffilmark{1},
K.~Hurley\altaffilmark{3},
J.~S.~Bloom\altaffilmark{1},
D.~A.~Frail\altaffilmark{4},
T.~J.~Galama\altaffilmark{1},
F.~A.~Harrison\altaffilmark{1},
G.~Morrison\altaffilmark{5},
D.~E.~Reichart\altaffilmark{1},
S.~A.~Yost\altaffilmark{1},
R.~Sari\altaffilmark{6},
T.~S.~Axelrod\altaffilmark{2},
T.~Cline\altaffilmark{7},
S.~Golenetskii\altaffilmark{8},
E.~Mazets\altaffilmark{8},
B.~P.~Schmidt\altaffilmark{2},
J.~Trombka\altaffilmark{7}.
}

\begin{abstract}
Of all the well localized gamma-ray bursts, GRB~000911 has the
longest duration ($T_{90}\sim 500\,$s), and ranks in the top 1\%
of BATSE bursts for fluence.  Here, we report
the discovery of the afterglow of this unique burst.
In order to simultaneously fit our radio and optical observations,
we are required to invoke a model involving an hard electron
distribution, $p \sim 1.5$ and a jet-break time less than 1.5~day.
A spectrum of the host galaxy taken 111 days after the burst reveals
a single emission line, interpreted as [OII] at a redshift $z = 1.0585$,
and a continuum break which we interpret as the Balmer limit at this
redshift.  Despite the long $T_{90}$, the afterglow of GRB 000911 is
not unusual in any other way when compared to the set of afterglows
studied to date.  We conclude that the duration of the GRB plays little
part in determining the physics of the afterglow.
\end{abstract}


\altaffiltext{1}{Palomar Observatory, 105-24, California Institute of
Technology, Pasadena, CA, 91125, USA.}
\altaffiltext{2}{Research School of Astronomy \& Astrophysics, Mount
Stromlo Observatory, Cotter Road, Weston, ACT, 2611, Australia.}
\altaffiltext{3}{University of California Space Sciences Laboratory, Berkeley, CA, 94720, USA.}
\altaffiltext{4}{National Radio Astronomy Observatory, P.O. Box O, Socorro, NM, 87801, USA.}
\altaffiltext{5}{Infrared Processing and Analysis Center 100-22, California
Institute of Technology, Pasadena, CA, 91125, USA.}
\altaffiltext{6}{Theoretical Astrophysics 130-33, California Institute of
Technology, Pasadena, CA, 91125, USA.}
\altaffiltext{7}{NASA/Goddard Space Flight Centre, Greenbelt, MD, 20771, USA.}
\altaffiltext{8}{Ioffe Physico-Technical Institute, St. Petersburg, 194021, Russia.}

\section{Introduction}

The revolution in understanding the origin of gamma-ray bursts (GRBs) has
been driven by the discovery and exploitation of long-lived lower-energy
``afterglow'' phenomenon (\cite{costa+97}; \cite{paradijs+97};
\cite{frail+97}).  However, this revolution has been restricted to GRBs
belonging to the so-called long/soft subclass (\cite{kouveliotou+93})
which are GRBs with duration, $T_{90} > 2$ s; here $T_{90}$ is the duration
of the 90th percentile of the fluence.  Technical difficulties have prevented
rapid and precise localisation of short bursts, and consequently these objects
continue to remain mysterious (see \cite{hurley+01b} for a summary of the
searches for afterglows of short-duration GRBs).

The currently popular theory for GRBs holds that the long/soft GRBs come
from the collapse of massive stars (\cite{mw99}), and 
short/hard GRBs are due to the merger of a pair of compact objects
(e.g. \cite{fryer+99}).  The former model naturally accounts for the
long duration since the newly formed black hole has access to a large
amount of stellar material whereas in the latter model the duration of
the accretion process is limited by the small amount of matter left
over following such a merger (e.g., \cite{janka+99}, \cite{npk01}).

Regardless of our current theoretical prejudice, the following three
phenomena give us the most intimate observational view of the central
engine: the variability, the duration of GRB profiles and the opening angles
of jets. In the ``Internal-External'' shock model for long-duration GRBs,
a model heavily supported by observations, the variability and the
duration have been argued to arise from different processes (\cite{np01}).  The recent
observational evidence for significant diversity in jet opening angles
(\cite{frail+01}) has motivated preliminary theoretical forays into the
same (\cite{vanP01}). 

Unfortunately, we are far from knowing how to translate these three
properties (variability, duration and opening angles) into physical
properties of the central engine.  In contrast, the study of the
afterglow yield good measures of bulk properties (explosion energy,
ambient density, location within the host galaxy, the host galaxy itself) 
as well as potentially direct information about the progenitors (X-ray
line features, late time red bumps potentially related to an underlying
supernova).  Thus the next natural step in GRB studies will be to start
empirically relating the engine properties to the afterglow properties.
It is in this context that the study of GRB~000911 with its unusually long
duration, $T_{90}\sim 500\,$s (the longest GRB with an identified afterglow
to date), becomes potentially interesting.  Here, we present the gamma-ray
time history, the discovery of the afterglow, broad-band modelling of the
afterglow and observations of the host galaxy of GRB~000911.

\section{Gamma-ray Observations}
\label{sec:GRB}

GRB~000911 was observed by three instruments in the Interplanetary
Network (IPN): Ulysses GRB (25--150 keV), Konus/Wind, and NEAR/XGRS
(100-1000 keV).  The earth-crossing time of the burst was 2000
September 11.30242 UT. By 22.8 hours after the burst, the event was
localized to an area of 30 arcmin$^{2}$ and reported promptly to the
GRB Coordinates Network (GCN)\footnote{\tt http://gcn.gsfc.nasa.gov}
(\cite{hurley+01a}).  

\begin{figure}
\centerline{\psfig{figure=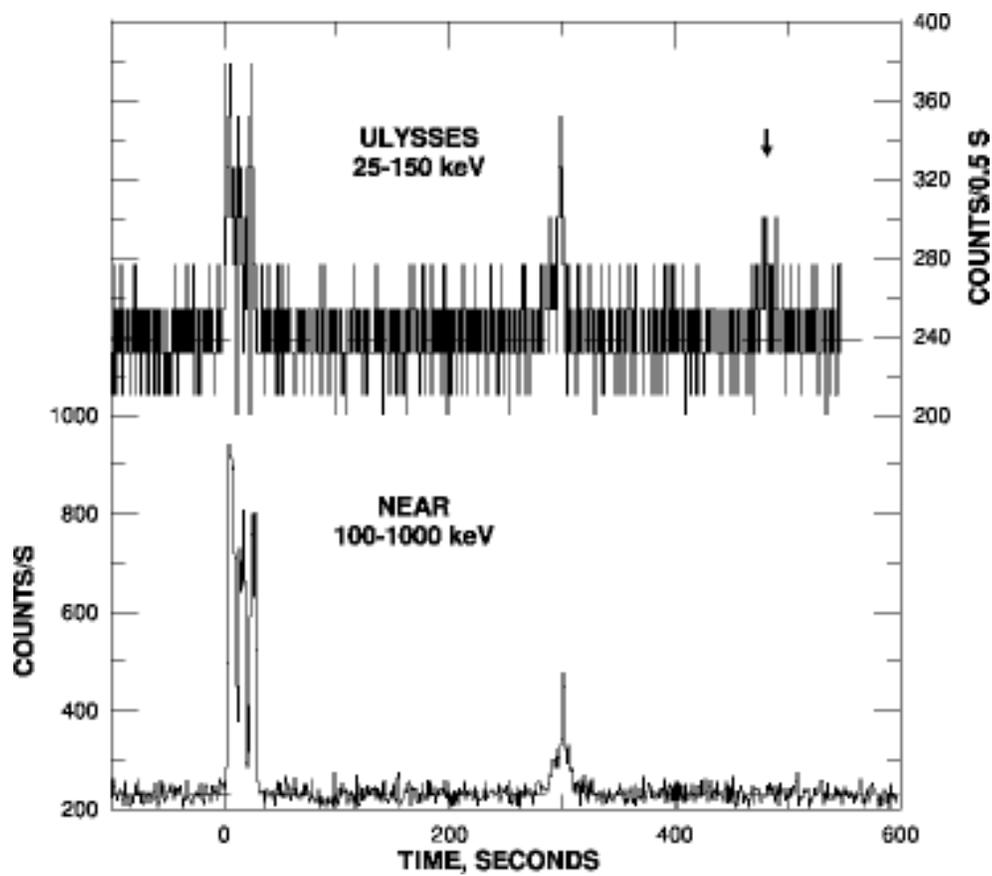,width=13cm,angle=-90}}
\caption{The gamma-ray light curves for GRB~000911 from Ulysees (top) and NEAR (bottom).}
\label{fig:Ulysses-NEAR-lightcurves}
\end{figure}

\begin{figure}[th]
\centerline{\psfig{figure=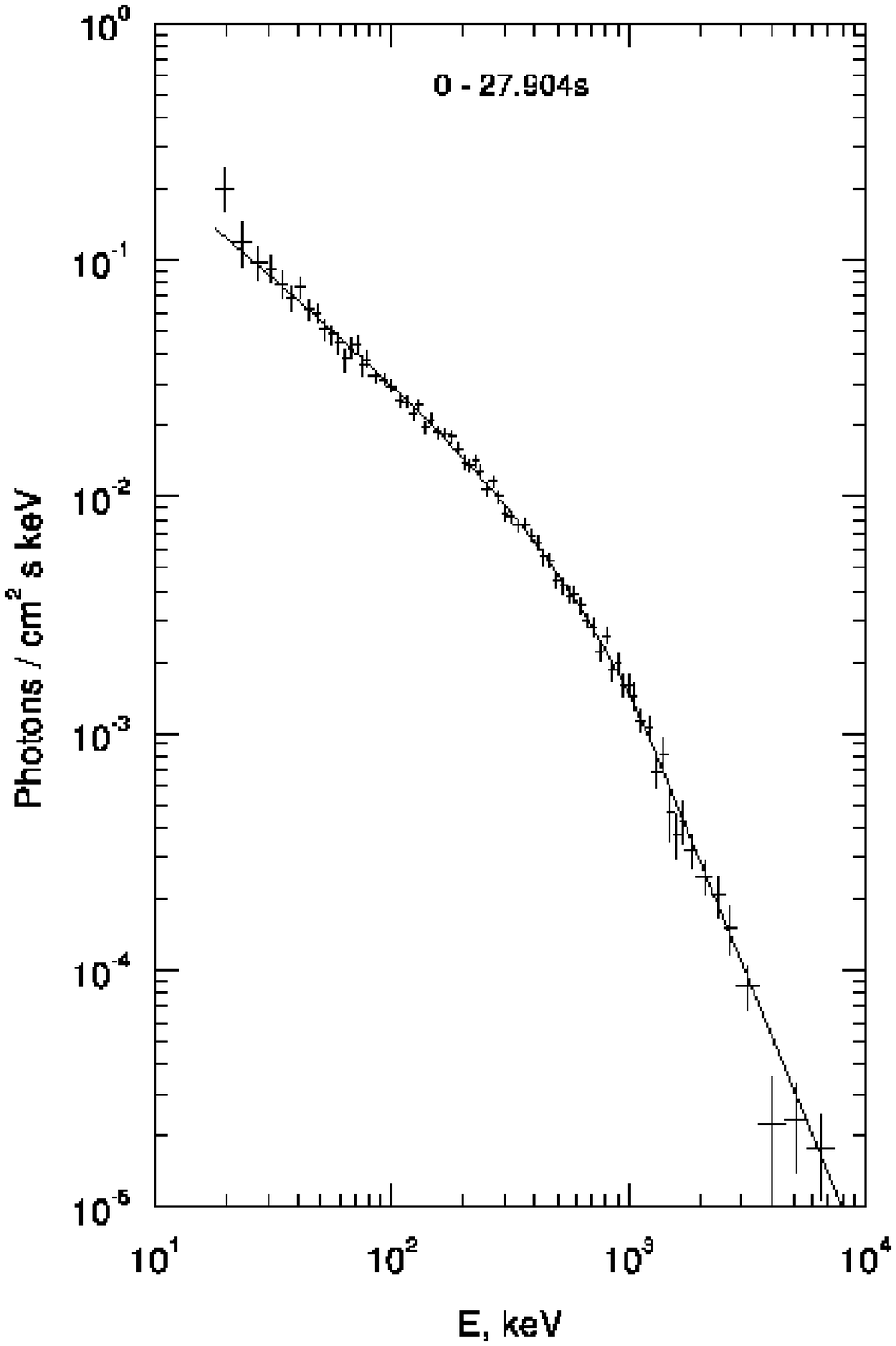,width=5cm}\qquad
            \psfig{figure=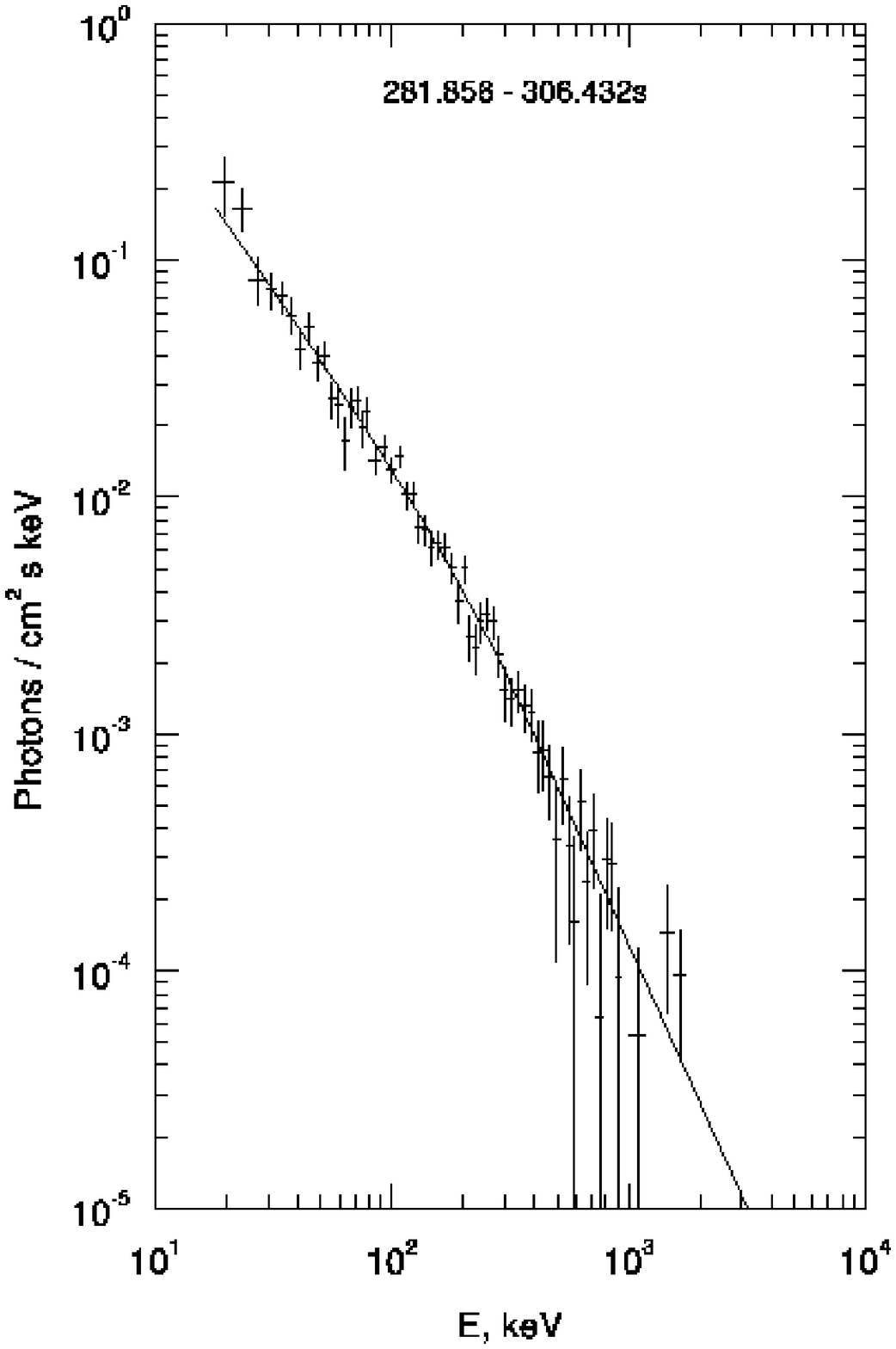,width=5cm}\qquad
            \psfig{figure=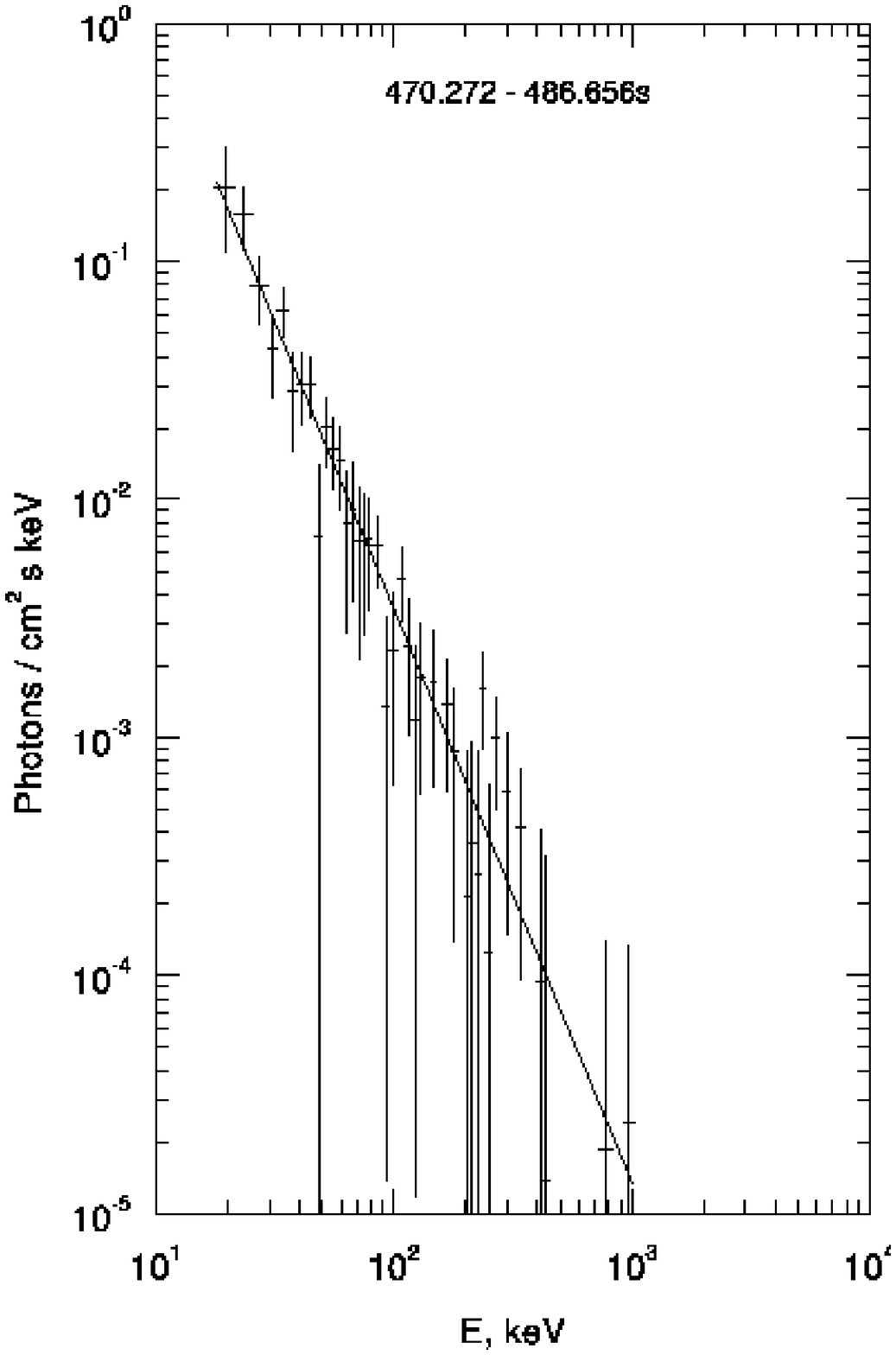,width=5cm}}
\caption[]{\small Konus/Wind spectra of of the three pulses
GRB 000911. The three pulses are indicated in Figure \ref{fig:Ulysses-NEAR-lightcurves}.
The epochs over which the spectra were extracted are displayed
inside the figure. The first two spectra were fitted to a Band model 
and the parameters displayed below the panels. There were not
enough counts in the third pulse and consequently the fit
was restricted to a simple power law model.}
\label{fig:Konus-spectra}
\end{figure}

\begin{figure}[tbp]
\centerline{\psfig{figure=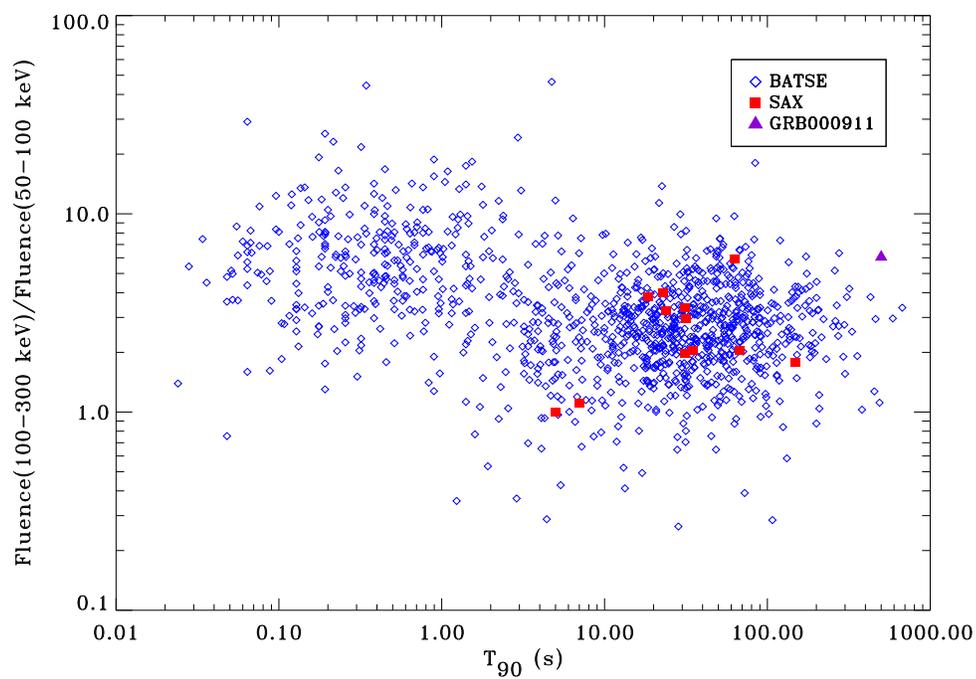,angle=90,width=13cm}}
\caption{The log $H$ -- log $T_{90}$ plot for 1676 BATSE GRBs (small
diamonds), demonstrating the two classes of GRBs (\cite{kouveliotou+93}).
Squares represent GRBs for which afterglows were identified with BeppoSAX.
It is clear that each belongs to the long/soft class.  The triangle (far
right) represents GRB~000911, which has the longest duration of all GRBs
with observed afterglows.}
\label{fig:2pop}
\end{figure}

Three distinct episodes (pulses) can be seen in the Ulysses light curve
and the first two episodes are clearly detectable in the NEAR data
(Figure~\ref{fig:Ulysses-NEAR-lightcurves}).  Energy spectra for each
pulse obtained from Konus/Wind (Figure~\ref{fig:Konus-spectra})
demonstrate the well-known hard-to-soft evolution and thereby support
the idea that all three pulses arise from the same GRB.  Furthermore,
given the Ulysses GRB detection rate of one per three days, we consider
it unlikely that the profiles are due to two distinct GRBs which are
coincidentally detected at about the same time.  Thus we conclude that
the duration of this burst is approximately 500\,s making it the longest
duration GRB whose afterglow has been studied to date (see
Figure~\ref{fig:2pop}).

The fluences of the three pulses are given in Table~\ref{tab:Fluences}.
With a total fluence of approximately $2.3 \times 10^{-4}$ erg cm$^{-2}$
this burst is within the top 1\% of the bursts observed by BATSE.
It is interesting to note that GRB 000911 is as bright as the famous
GRB~990123 (\cite{bbk+99}).

\section{Discovery of the Optical and Radio Afterglow}

The IPN error box was first observed by the robotic 50-inch telescope
at Mount Stromlo Observatory with the MACHO dual-beam camera 1.4 days
after the burst.  
Very Large Array (VLA\footnotemark\footnotetext{The National Radio
Astronomy Observatory (NRAO) is a facility of the National Science
Foundation operated under cooperative agreement by Associated Universities,
Inc.  NRAO operates the VLA.}) observations of GRB~000911 were initiated
0.8 days later at 8.46 GHz.  All radio observations were obtained
in the standard continuum mode with $2\times 50$ MHz contiguous bands.  
We observed the extra-galactic source J0203+115 for phase calibration and 3C48
(J0137+331) for flux calibration.  All data were reduced using the
Astronomical Image Processing System (AIPS).  

\begin{figure}[tbp]
\centerline{\psfig{figure=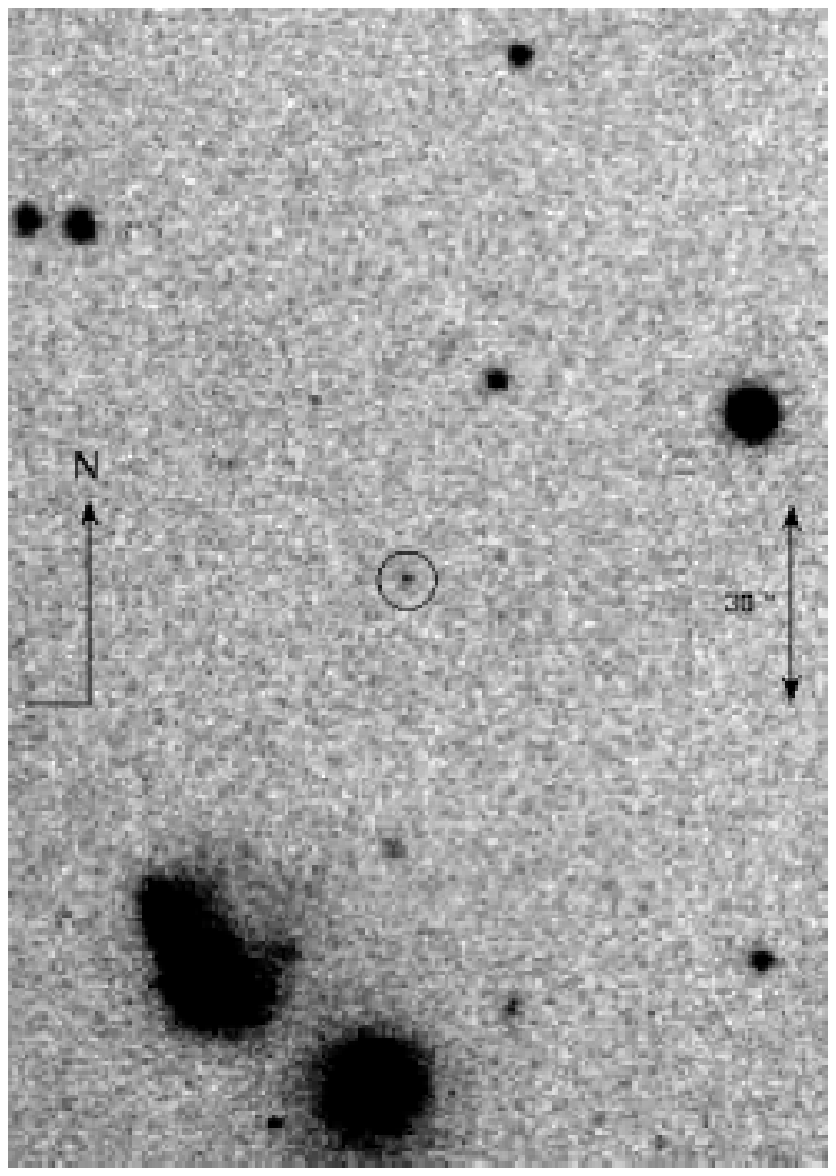,width=6cm}\psfig{figure=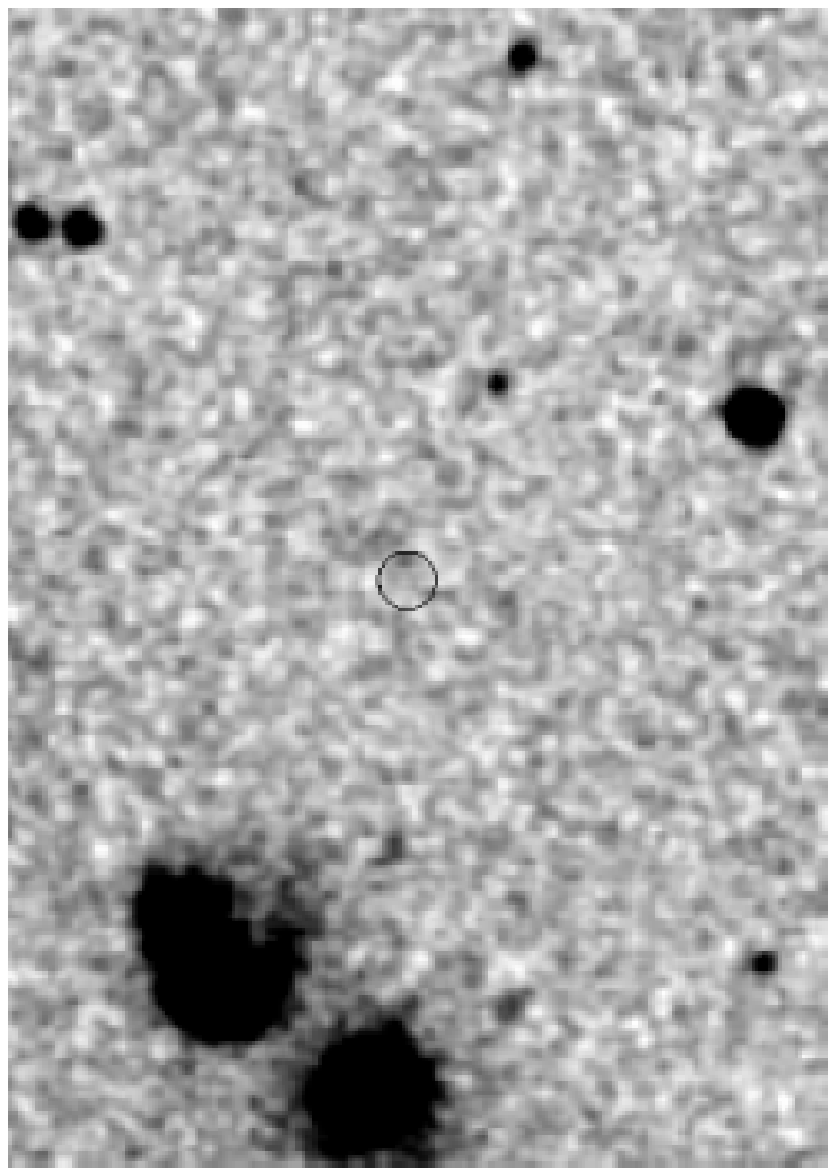,width=6cm}}
\caption{Identification of the optical afterglow of GRB~000911.  Left: $R_{M}$
image from the 50-inch telescope at MSO.  Right: Corresponding image from
the DPOSS $F$-plate.  The location of the optical afterglow is circled.}
\label{fig:MSO}
\end{figure}

In the earliest radio observations (2.2 and 3.1 days after the burst), we
identified as a candidate afterglow a source located at J2000 coordinates
02:18:34.36 +07:44:27.65 (with an estimated 2 $\sigma$ error of 0.6
arcseconds in each axis) that brightened by 2 $\sigma$.   An optical
counterpart to this radio source (Figure~\ref{fig:MSO}) with
$R\sim 20.6$ mag was identified in our images obtained from
the 50-inch MSO telescope but the source was not
present on the Digital Palomar Observatory Sky Survey (\cite{djorgovski+99}),
although this is near the plate limit.  Subsequent observations with
Echellette Spectrograph and Imager (ESI) on the Keck II telescope of
the W.~M.~Keck Observatory showed that the source decayed and thereby
confirmed the identification of this source as the 
afterglow of GRB~000911. 

Subsequently we began a program of VLA observations of the radio
afterglow. The log of radio observations and the fluxes of the
radio afterglow can be found in 
Table~\ref{tab:radiodata}.

We observed the optical afterglow at the Keck Observatory in
$R$ and $I$ bands at three additional epochs after the discovery.
We bias-subtracted and flat-fielded the optical observations
in the usual way, using the Image Reduction and Analysis Facility
(IRAF)\footnote{IRAF is distributed by the National Optical Astronomy
Observatories.}.  Aperture photometry was used to measure the magnitude
of the optical afterglow in each image.  Photometric zero points for
the Keck imaging were based on stars in the field, with calibrated values
provided by S.~Covino and D.~Lazzati from their first VLT epoch
(\cite{lazzati+01}).  In addition, the large range of the color of
stars in the field [$(R-I) \sim$ 0.4 to 1.8 mag] allowed us to fit
and apply colour terms.

The first epoch optical 
images (in MACHO blue and red bands) are not sufficiently
deep to use the calibrated VLT stars.  For these images, we transform
the stellar magnitudes calibrated by Henden (2000)\nocite{henden00} to the MACHO
photometric system using (the inverse of) the transformation calculated by
Bessell \& Germany (1999)\nocite{bg99}.  We then photometered the afterglow
against these stars and transformed back to the Johnson-Cousins system.
An additional error of 3\% was added in quadrature to these measurements
to reflect errors in the transformation.

The log of our optical observations and the photometry are displayed in
Table \ref{tab:optdata}.  All the optical data, photometric and
spectroscopic, when displayed in physical units (Jy), have been
corrected for foreground Galactic reddening $E_{B-V} = 0.107$ mag
(\cite{sfd98}) and an average Galactic extinction curve (with $R_{V} = 3.1$)
modelled by Cardelli, Clayton, \& Mathis (1989)\nocite{ccm89}.

\begin{figure}[tbp]
\centerline{\psfig{figure=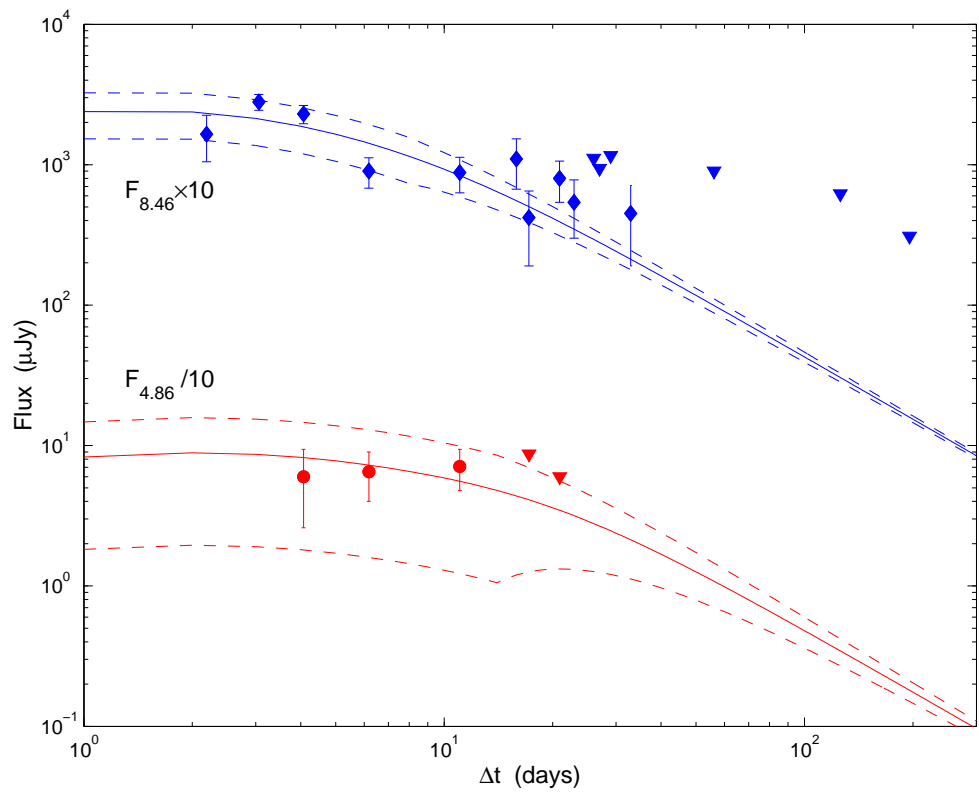,width=13cm}}
\caption{Results of the fit to the radio measurements of the afterglow of
GRB~000911.  The solid line indicates our fit afterglow model light-curve,
while the dashed lines indicate the scintillation envelope.}
\label{fig:LC-radio}
\end{figure}

\begin{figure}[tbp]
\centerline{\psfig{figure=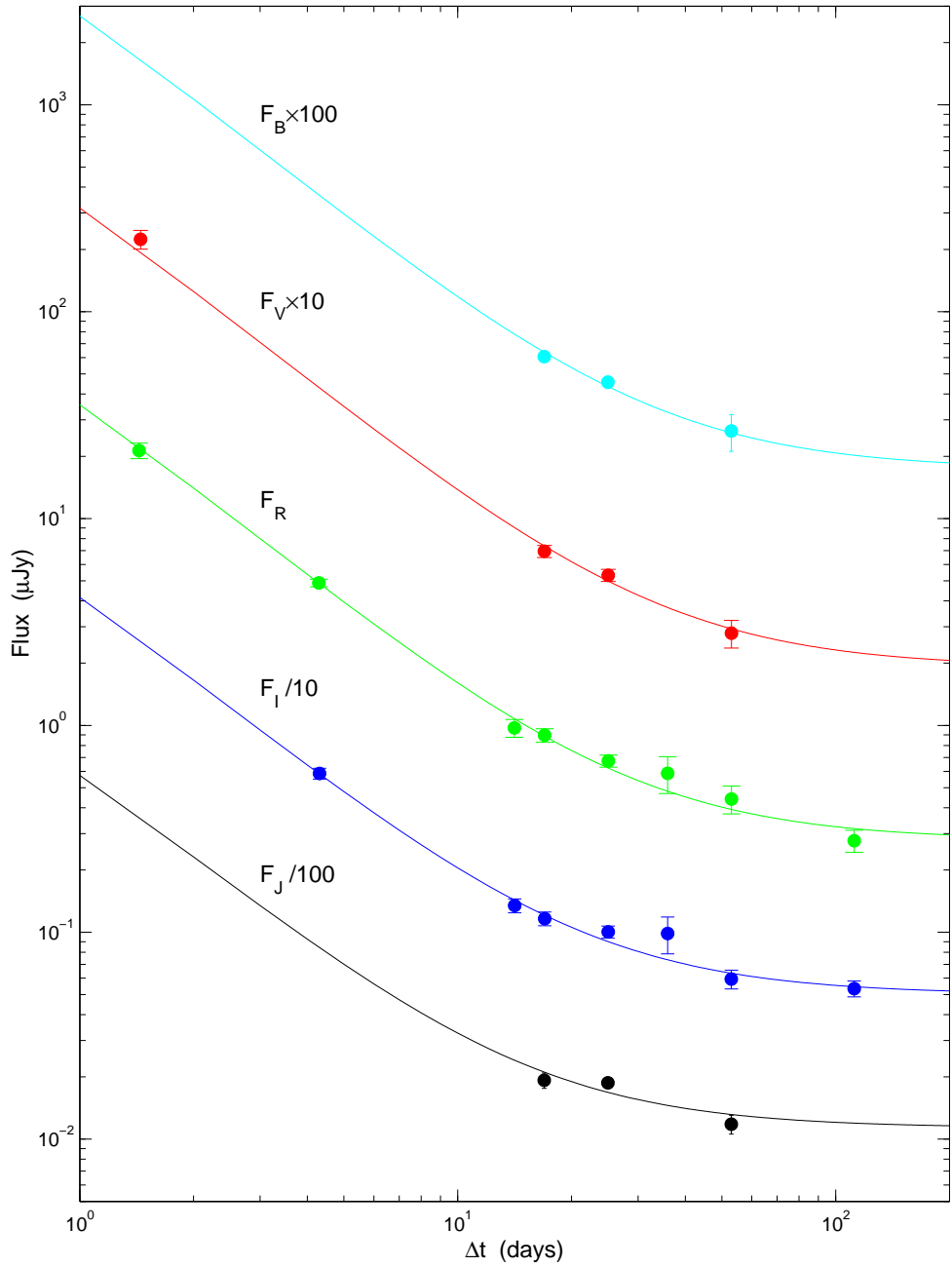,width=13cm}}
\caption{Results of the fit to our optical measurements of the afterglow of
GRB~000911, and those from Lazzati et al. (2001).  The data displayed here
have been corrected for Galactic extinction.}
\label{fig:LC-optical}
\end{figure}

Optical and radio light curves are displayed in Figures~\ref{fig:LC-radio}
and \ref{fig:LC-optical} respectively. The optical light-curves (which
include data from from Lazzati et al. (2001)\nocite{lazzati+01})
 exhibit power law decay, $f(t)\propto t^{-\alpha}$ with $\alpha=1.46
\pm 0.05$, followed by a levelling (at about 15 days).  We identified
the late time steady source with the host galaxy of GRB~000911.  

\section{The Afterglow Model}
\label{sec:afterglow-model}

We modeled the radio and optical light curves using broad-band
modeling programs that have been used for other GRBs (see, e.g.,
\cite{berger+00},  
\cite{pk01}).

The basic afterglow model assumes a relativistic spherical expansion
into a constant-density (\cite{spn98}; hereafter ISM) or wind-shaped
medium (\cite{cl00}; hereafter WIND).  In this model electrons are
accelerated at the shock front into a power law energy distribution,
$N(E)\propto E^{-p}$, with $p \sim 2.4$, and emit synchrotron radiation
with a spectrum described by three break frequencies and an overall
flux normalization, $F_{m}$ at the synchrotron peak frequency
$\nu_{m}$.  The temporal evolution of the spectrum depends on the
density, $n$ and density structure (ISM or WIND) of the ambient
environment.  In addition to $p$ the other microphysics parameters are
the electron equipartition factor $\epsilon_{e}$, and magnetic field
equipartition factor $\epsilon_{B}$.

The ISM model fails to simultaneously fit both the optical and radio
data since in this model $F_{m}$ remains constant with time.  As a result,
this model predicts a lower flux density level than what is observed in
the radio regime.  This excess emission is too strong and consistently
high to be explained by interstellar scintillation.  Explanation in terms
of a reverse shock contribution in the radio at early times (e.g.
\cite{frail+00}) also proved to yield unsatisfactory results, with a
$\chi^2/{\rm DOF} = 40/31$.  

For the WIND model we again fail to find a satisfactory solution, with
our best $\chi^{2}=52$ for 32 DOF (we were unable to constrain the cooling
frequency in a meaningful way, and so it was fixed).  In this case we encounter a similar
problem in which the model predicts a lower flux level than is observed
in the radio at early times.

We then considered the ISM model but allowed for a conical geometry
(jets) for the explosion (\cite{sph99}).  In this model (hereafter
ISM+Jet), the jet edge becomes visible to observers at epoch ($t_j$)
when the bulk Lorentz factor $\Gamma$ falls below $\theta_{\rm jet}^{-1}$,
the inverse of the jet opening angle.  Prior to this jet break time the
evolution of the synchrotron spectrum is equivalent to the spherical
evolution scenarios.  For $t\age t_j$ the lightcurves are expected to
be:  $F_{\nu} \propto t^{-1/3}$ or $F_{\nu} \propto t^{-p}$ for $\nu <
\nu_{m}$ and $\nu > \nu_{m}$, respectively.

The  best fit parameters for the jet model are: $t_j=0.6\,$d,
$\nu_a=20.5\,$GHz, $\nu_m=460\,$GHz, $\nu_c=8\times 10^{3}\,$GHz,
$f_m,=4.4\,$mJy and $p=1.49$; here the epoch for the frequencies is
$t_j$.  Thus the ISM+Jet model requires a very ``flat $p$'' ($p < 2$).
Such flat $p$ values have been inferred in the afterglows of several
other GRBs, e.g. GRB 000301C, GRB 010222 (see \cite{pk01}).

Afterglow with $p<2$ must have a high frequency cutoff so that the
total energy in the afterglow does not diverge.  Bhattacharya
(2001)\nocite{bhattacharya01} has shown that the temporal evolution
of the spectrum of an afterglow with $p < 2$ is identical to that for
$p > 2$ provided the maximum Lorentz factor of the electron distribution
scales linearly with the Lorentz factor of the shock.  This requirement
has yet to be demonstrated, but appears to be a reasonable assumption.
Accepting this {\it ad hoc} requirement, we can convert the above inferred
parameters to yield the following physical parameters:
$p = 1.5$, $E_{52} = 1.1$, $\epsilon_{B} = 2.3$, $\epsilon_{e} = 0.016$
and $n = 0.07$ cm$^{-3}$.  We warn the reader that given the paucity of the
afterglow data, each of these values, except for $p$, are subject to the
large errors in the fit break frequencies and should not be taken as
definitive (the error in $p$ is set by the error in the optical decay slope,
which is small).  The unphysical value of $\epsilon_B$ ($>1$) is likely
the result of these large errors.  The overall $\chi^2$/DOF=29.7/31 is
quite good and the results of the fits can be found in
Figures~\ref{fig:LC-optical} and \ref{fig:LC-radio}.

\section{Host Galaxy}
\label{sec:host}

Spectra of the host galaxy were obtained on Keck II 01 January 2001 UT,
in excellent conditions, using the Low-Resolution Imaging Spectrometer
(LRIS; \cite{occ+95}).  We used a 1.0-arcsec wide long-slit, at a
position angle $155.1^\circ$, close to the parallactic, and a 300 lines
mm$^{-1}$ grating, giving an effective instrumental resolution FWHM
$\approx 12\,$\AA, and an approximate wavelength coverage 3950--8900 \AA.
We obtained 4 exposures of 1800 s each, with small dithers on the slit.  
Exposures of internal flat-field lamp and arc lamps were obtained for
calibrations.  Small wavelength shifts due to the instrument flexure
were removed by fitting the wavelengths of strong, unblended night sky 
lines.  The net resulting wavelength scatter is $\sim 0.2$ \AA. 
Since no adequate flux calibration was obtained for this night, we used the
response curves determined from exposures of standard stars 
BD$+17^\circ 4708$ and BD$+26^\circ 2606$ from Oke \& Gunn (1983)\nocite{og83},
obtained 2 nights before with the same instrument setup.  We removed the
residual flux zero-point uncertainty (due to the differential slit losses
and seeing variations) by matching the spectroscopic $BVR$ magnitudes to
those determined photometrically; we estimate the net flux zero-point
uncertainty to be $\sim 20$\%.

\begin{figure}[tbp]
\centerline{\psfig{figure=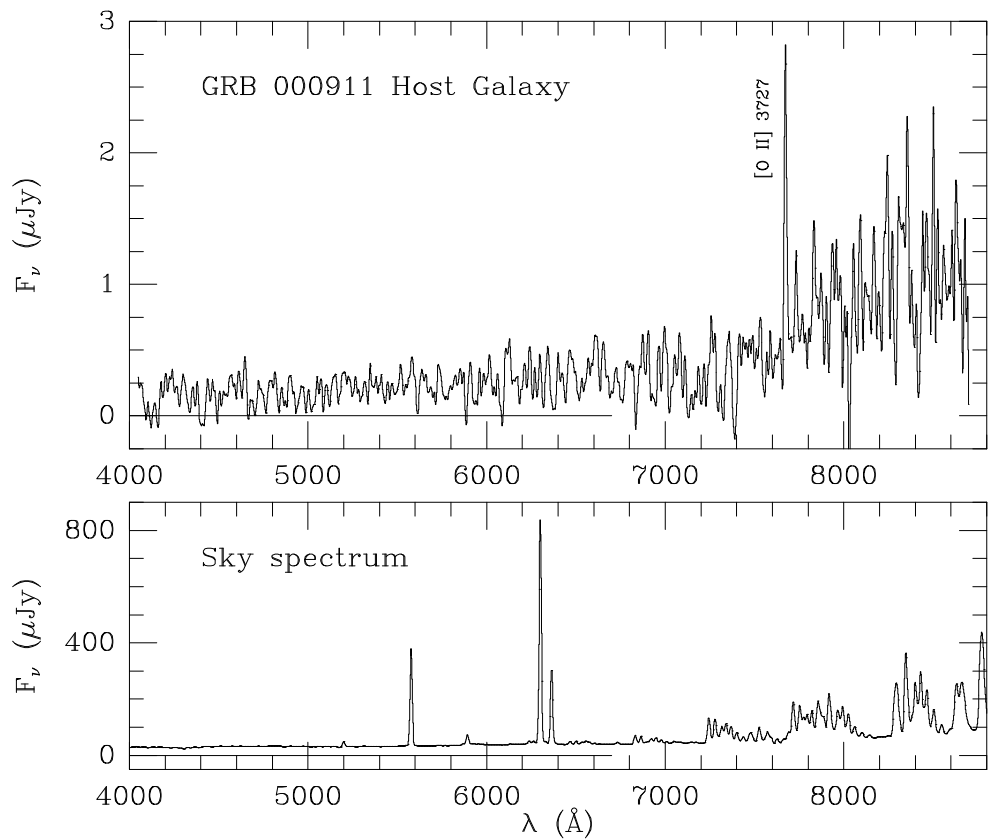,width=13cm}}
\caption{The Keck spectrum of the host galaxy of GRB 000911 (top), and the
corresponding night sky spectrum (bottom).  The spectra have been smoothed
with a Gaussian with $\sigma = 5$ \AA, corresponding to the instrumental
resolution.  The [O II] 3727 emission line is labeled; all other apparent
spikes are due to imperfect subtraction of the strong night sky lines.
The observed continuum drop near 7500 \AA\ corresponds to the Balmer break
at the redshift of the object, $z = 1.0585$.}
\label{fig-spectrum}
\end{figure}

The final combined spectrum of the host galaxy is shown in
figure~\ref{fig-spectrum}.  
A strong emission line is detected at $\lambda_{obs} = 7673.1 \pm 0.3$ \AA.
Given the presence of a well-detected continuum blueward of the line, and
the absence of other strong lines, we interpret this line as the commonly
detected [O II] 3727 doublet at a redshift $z = 1.0585 \pm 0.0001$.  Note
that the instrumental resolution was not high enough to resolve this line
as a doublet.  There is also a hint of the possible H$\delta$ line at 
$\lambda_{obs} \approx 8442$ \AA, but the difficulty inherent in subtracting
the strong night sky lines in this part of the spectrum precludes a definitive
measurement of this or other high-order Balmer lines.
However, there is a clear continuum decrement corresponding to the
Balmer limit at this redshift, and thus we consider our redshift interpretation
to be secure.

In what follows, we assume for the foreground Galactic reddening
$E_{B-V} = 0.107$ mag (\cite{sfd98}) and a standard Galactic extinction
curve (\cite{ccm89}). The fluxes quoted are uncertain by about 20\%,
with contributions  from the statistical noise and the systematic flux
zero-point uncertainty contributing about 10\% each.

The observed [O II] 3727 line 
flux\footnotemark\footnotetext{The fluxes quoted are uncertain by about 20\%,
with contributions  from the statistical noise and the systematic flux
zero-point uncertainty contributing about 10\% each.} is $(1.9 \pm 0.2)
\times 10^{-17}$ erg cm$^{-2}$ s$^{-1}$.  Correcting for the Galactic
extinction, we obtain $f_{3727} = (2.3 \pm 0.3) \times 10^{-17}$ erg
cm$^{-2}$ s$^{-1}$.  The observed equivalent width is $W_\lambda = 63
\pm 3$ \AA, i.e., $30.5 \pm 1.5$ \AA\ in the restframe.
This is typical for field galaxies at comparable redshifts, which have
restframe equivalent widths spanning 0 -- 50 \AA\ (\cite{hcbp98}).
The observed continuum flux near $\lambda \approx 5764$ \AA, corresponding to
the restframe wavelength of 2800 \AA, is $\approx 0.2$ $\mu$Jy.
Correcting for the Galactic extinction, we obtain $f_{2800} \approx
0.26$ $\mu$Jy.  Finally, from the observed $I$ band magnitude of 24.3 mag
the host, we estimate the observed flux corresponding to the restframe $B$ band
to be  $\approx 0.45$ $\mu$Jy.  Correcting for the extinction, we obtain
$f_{B} \approx 0.53$ $\mu$Jy.

We will assume a flat cosmology with $H_0 = 65$ km s$^{-1}$
Mpc$^{-1}$, $\Omega_M = 0.3$, and $\Omega_{\Lambda} = 0.7$.  For $z = 1.0585$, 
the luminosity distance is $2.355 \times 10^{28}$ cm, 
and 1 arcsec corresponds to 8.7 proper kpc in projection.
From the [O II] 3727 line flux, we calculate the line luminosity
$L_{3727} = 1.6 \times 10^{41}$ erg s$^{-1}$.  Using the star formation
rate estimator from Kennicutt (1998)\nocite{kennicutt98},  we derive the star
formation rate  $SFR_{3727} \approx 2.2~M_\odot$ yr$^{-1}$.
From the UV continuum power at $\lambda_{rest} = 2800$ \AA, using the
estimator from Madau, Pozzetti \& Dickinson (1998)\nocite{mpd98}, we obtain
$SFR_{2800} \approx 1.1~M_\odot$ yr$^{-1}$.  The relatively small difference
between these two estimates may be due in part to a combination of our
measurement errors (which are about 20\% for each value), the intrinsic
scatter of these estimators (which is at least 30\%), and possibly to some
reddening in the host galaxy (the flux at $\lambda_{rest} = 2800$ \AA\ would
be extincted more than the flux at $\lambda_{rest} = 3727$ \AA).
Our optically derived SFR is a lower limit since the optical/UV tracers
are insensitive  to star-formation arising in dusty regions.

From the observed continuum flux corresponding to the restframe $B$ band
we derive $M_B \approx -18.95$ mag, i.e. the luminosity (then) corresponding
to $\sim 0.3 ~L_*$ (today).  While the interpretation of this measurement
hinges on the assumed evolution of the object (i.e. it may be a progenitor
of a present-day spiral, or a  dwarf galaxy), we note that it is comparable
to the observed luminosities of other GRB hosts (\cite{hf99}).

We measure an offset from the afterglow to the host of $0.08 \pm 0.10$ arcsec,
well within the marginally resolved host (FWHM $\sim$ 720 mas) and consistent
with the offsets for other GRBs found by Bloom, Kulkarni \& Djorgovski (2000)
\nocite{bkd00}.  Thus even in this respect we find that GRB~000911 appears
to be no different from other long/soft GRBs.

\section{Conclusions}
\label{sec:conclusions}

With a duration of 500 s, GRB~000911 is the longest duration burst to be
well localized to date.  Here we have presented the high energy profiles
and the evolution of the spectrum.  Thanks to a combination of prompt radio
and optical observations we were able to discover and study the afterglow.
After the afterglow had faded, we observed the host galaxy and obtained
its redshift, $z=1.0585$.

We obtained a modest amount of optical and radio data of the
afterglow.  The best fit model requires a flat electron energy index,
$p\sim 1.5$ and a break in the afterglow at about $\sim 0.6$ d after
the burst.  Following the usual interpretation, we attribute this break
to a jet.  At a redshift of 1.0585, the isotropic $\gamma$-ray energy
release (using the Konus fluence) is $7.8 \times 10^{53}$ erg.  Using
the particular formulation given in Frail et al. (2001)\nocite{frail+01}
the jet break time of $0.6\,$d translates to a jet opening angle,
$\theta_j=1.9^\circ$.  The $\gamma$-ray energy corrected for the conical
geometry is $E_\gamma\sim 3 \times 10^{50} n_{0.1}^{1/4}$ erg (\cite{sph99})
where $n_{0.1}$ is the ambient density in units of 0.1 cm$^{-3}$.  This
value is consistent with the median $<E_{\gamma}>$ found by Frail et al.
(2001)\nocite{frail+01} of $5 \times 10^{50}$ erg with a $1\sigma$ spread
of a factor of two.  Since the jet break must have occurred prior to our
first observation, a more conservative value of $t_{\rm j} < 1.5$ d yields
an upper limit on the jet-corrected energy release of $\ale 4.2 \times
10^{50} n_{0.1}^{1/4}$ erg, again consistent with the result of Frail et
al.(2001).

The properties of this afterglow is quite similar to that 
of the short duration burst ($T_{90} \sim 2$ s) GRB 000301C
(\cite{panaitescu01}).  Both are flat spectrum bursts and with a
jet around a day or so.  Thus it appears that the duration of the
burst had little effect on either the jet opening angle or on the
afterglow parameters.

Following the discovery of afterglow of GRB~000911, Lazzati et al.
(2001)\nocite{lazzati+01} organized a program at ESO and Keck to follow up
the optical afterglow at late times and found marginal evidence for an
underlying supernova (SN).  Such late time red-bumps were first noted in
GRB~980326 (\cite{bkd+99}) and GRB~970228 (\cite{reichart99,gtv+00}) and
interpreted as underlying SN. Other authors have attributed such bumps to
dust echoes (\cite{eb00}).  In any case, the presence of such late time red
bumps seem to occur in GRBs with different durations. Finally, the host galaxy
of GRB 000911 appears to be quite representative of host galaxies.

We conclude that there is no compelling reason to believe that
the duration of GRBs (when restricted to the sub-class of
long duration events) plays a significant role in determining
either the afterglow properties and perhaps even particularly
are related to the properties of the central engine (the jet opening
angle).

\section*{Acknowledgements}

SRK and SGD thank NSF for support of our ground-based GRB observing program.
We are grateful to the staff of MSSSO and Keck observatories for their
expert help.  PAP gratefully acknowledges an Alex Rodgers' Travelling
Scholarship.  JSB gratefully acknowledges the fellowship support from the
Fannie and John Hertz Foundation.  KH is grateful for Ulysses support under
JPL Contract 958056, and for IPN support under the NEAR Participating
Scientist program, NAG5-9503, and under the LTSA, NAG5-3500.

\clearpage

\begin{deluxetable}{cccc}
\footnotesize
\tablecolumns{4}
\tablewidth{0pc}
\tablecaption{Fluences of the Three Pulses}
\tablehead{\colhead{Epoch (s)} & \colhead{Fluence} &
        \colhead{Peak Flux} & \colhead{Energy Interval}}
\startdata
(s)           &    erg cm$^{-2}$  & erg cm$^{-2}$ s$^{-1}$ & keV \\
0--27.9       & $2.0\times 10^{-4}$ & $2\times 10^{-5}$   & 15--8000 \\
281.9-306.4   & $2.7\times 10^{-5}$ & --                & 15--5000 \\
470.3--486.7  & $4.0\times 10^{-6}$ & --                & 15-1000 \\
\enddata
\tablecomments{Peak fluxes for the second and third pulses are not
shown because the time bins were too large, 8.192 s.}
\label{tab:Fluences}
\end{deluxetable}

\clearpage

\begin{deluxetable}{lcc}
\tabcolsep0in\footnotesize
\tablewidth{\hsize}
\tablecaption{Radio Observations of GRB~000911 made with the Very Large Array.\label{tab:radiodata}}
\tablehead {
\colhead {Epoch}      &
\colhead {$\nu_0$} &
\colhead {S$\pm\sigma$} \\
\colhead {(UT)}      &
\colhead {(GHz)} &
\colhead {($\mu$Jy)}
}
\startdata
2000 Sep 13.49 & 8.46 & 165$\pm$60    \nl
2000 Sep 14.36 & 8.46 & 278$\pm$36    \nl
2000 Sep 15.36 & 4.86 & 60$\pm$34    \nl
2000 Sep 15.36 & 8.46 & 230$\pm$34    \nl
2000 Sep 17.49 & 4.86 & 65$\pm$25    \nl
2000 Sep 17.49 & 8.46 & 90$\pm$22    \nl
2000 Sep 22.35 & 4.86 & 71$\pm$23    \nl
2000 Sep 22.35 & 8.46 & 88$\pm$25    \nl
2000 Sep 28.50 & 4.86 & 27$\pm$31    \nl
2000 Sep 28.50 & 8.46 & 42$\pm$23    \nl
2000 Oct 2.22  & 4.86 & 10$\pm$25    \nl
2000 Oct 2.22  & 8.46 & 80$\pm$26    \nl
2000 Oct 4.23  & 8.46 & 54$\pm$24    \nl
2000 Oct 8.29  & 8.46 & -6$\pm$50    \nl
2000 Oct 10.27 & 8.46 & -46$\pm$81    \nl
2000 Oct 14.24 & 8.46 & 45$\pm$26    \nl
2000 Nov 7.38  & 8.46 & 30$\pm$30    \nl
2001 Jan 15.08 & 8.46 & 26$\pm$18    \nl
\enddata
\tablecomments{The columns are (left to right), (1) UT date of the
  start of each observation, (2) time elapsed since the $\gamma$-ray
  burst, (3) telescope name, (4) observing frequency, and (5) peak
  flux density at the best fit position of the radio transient, with
  the error given as the root mean square noise on the image.}
\end{deluxetable}

\clearpage

\begin{deluxetable}{cccc}
\footnotesize
\tablecolumns{4}
\tablewidth{0pc}
\tablecaption{\label{tab:optdata}Optical observations of the afterglow of
GRB~000911.  These measurements have not been corrected for Galactic
extinction.}
\tablehead{\colhead{Date (UT)} & \colhead{Filter} & \colhead{Magnitude} & \colhead{Telescope}}
\startdata
2000 Sep 12.737	&	V   &	20.90   $\pm$	0.10	&	MSO50		\\
2000 Sep 12.737	&	R   &	20.700  $\pm$	0.081	&	MSO50		\\
2000 Sep 15.593	&	R   &	22.299  $\pm$	0.030	&	Keck II + ESI	\\
2000 Sep 15.606	&	I   &	21.816  $\pm$	0.052	&	Keck II + ESI	\\
2000 Sep 25.421	&	R   &	24.052  $\pm$	0.094	&	Keck I + LRIS	\\
2000 Sep 25.451	&	I   &	23.412  $\pm$	0.071	&	Keck I + LRIS	\\
2001 Jan 1.221	&	R   &	25.414  $\pm$	0.119	&	Keck II + ESI	\\
2001 Jan 1.248	&	I   &	24.415  $\pm$	0.083	&	Keck II + ESI	\\
\enddata
\end{deluxetable}

\clearpage

\clearpage

\clearpage

\clearpage

\clearpage

\clearpage

\clearpage

\clearpage


\begin{thebibliography}{}
\bibitem[Berger et~al. 2000]{berger+00}
Berger, E.J., et~al., 2000, ApJ, 545, 56.

\bibitem[Bessell \& Germany 1999]{bg99}
Bessell, M.S. \& Germany, L.M., 1999, PASP, 111, 1421.

\bibitem[Bhattacharya 2001]{bhattacharya01}
Bhattacharya, D., 2001, BASI (submitted), astro-ph/0104250.

\bibitem[Esin \&\ Blandford 2000]{eb00}
Esin, A. A. \&\ Blandford, R. D. 2000, ApJ 534, L151 


\bibitem[Bloom, Kulkarni \& Djorgovski 2000]{bkd00}
Bloom, J.S., Kulkarni, S.R. \& Djorgovski, S.G., 2000, AJ (submitted),
astro-ph/0010176.

\bibitem[Bloom et~al. 1999]{bkd+99}
Bloom, J.S., et~al., 1999, Nature, 401, 453.

\bibitem[Briggs et al. 1999]{bbk+99}
Briggs, M. S. et al. 1999, ApJ 524, 82.

\bibitem[Cardelli, Clayton \& Mathis 1989]{ccm89}
Cardelli, J.A., Clayton, G.C. \& Mathis, J.S., 1989, ApJ, 345, 245.


\bibitem[Chevalier \& Li 2000]{cl00}
Chevalier, R.A. \& Li, Z.-Y., 2000, ApJ, 536, 195.

\bibitem[Costa et al. 1997]{costa+97}
Costa, E., et al., 1997, Nature, 387, 783.

\bibitem[Djorgovski et~al. 1999]{djorgovski+99}
Djorgovski, S.~G., Gal, R.~R., Odewahn, S.~C., DeCarvalho, R.~R., Brunner, R.,
Longo, G. \& Scaramella, R. 1999, in {\it Wide Field Surveys in Cosmology},
S. Colombi et al., eds., p. 89.

\bibitem[Frail et~al. 2001]{frail+01}
Frail, D.A., et~al., 2001, astro-ph/0102282.

\bibitem[Frail et~al. 2000]{frail+00}
Frail, D.A., et al., 2000, ApJ, 538, L129.

\bibitem[Frail et al. 1997]{frail+97}
Frail, D.A., Kulkarni, S.R., Nicastro, S.R., Feroci, M. \& Taylor, G.B., 1997, Nature, 389, 261.

\bibitem[Frontera et al. 2000]{frontera+00}
Frontera, F., et al., 2000, ApJSS, 127, 59.

\bibitem[Fryer et al. 1999]{fryer+99}
Fryer, C.L., Woosley, S.E., Herant,M., \& Davies, M.B., 1999, ApJ, 520, 650.

\bibitem[Galama et al. 2000]{gtv+00}
Galama, T. et al. 2000, ApJ 536, 185.

\bibitem[Galama et~al. 2001]{galama+01}
Galama, T.J., et al., 2001, in preparation.

\bibitem[Henden 2000]{henden00}
Henden, A., 2000, GCN Circular 800.

\bibitem[Harrison et~al. 2001]{harrison+01}
Harrison, F.A., et al., 2001, ApJ (submitted), astro-ph/0103377.


\bibitem[Hogg \& Fruchter 1999]{hf99}
Hogg, D.W. \& Fruchter, A.S., 1999, ApJ, 520, 54.

\bibitem[Hogg et al. 1998]{hcbp98}
Hogg, D., Cohen, J., Blandford, R., \& Pahre, M. 1998, ApJ, 504, 622.

\bibitem[Hurley et al. 2001a]{hurley+01a}
Hurley, K., et al., 2001a, GCN Circular 791.

\bibitem[Hurley et al. 2001b]{hurley+01b}
Hurley, K., et al., 2001b, ApJ (submitted), astro-ph/0107188.

\bibitem[Janka et al. 1999]{janka+99}
Janka, H.-T., Eberl, T., Ruffert, M., \& Fryer, C.L., 1999, ApJ, 527, L39.

\bibitem[Kennicutt 1998]{kennicutt98}
Kennicutt, R.C., 1998, ARA\&A, 36, 189.

\bibitem[Kouveliotou et al. 1993]{kouveliotou+93}
Kouveliotou, C., Meegan, C.A., Fishman, G.J., Bhat, N.P., Briggs, M.S.,
Koshut, T.M., Paciesas, W.S. \& Pendleton, G.N., 1993, ApJ, 413, L101.

\bibitem[Lazzati et~al. 2001]{lazzati+01}
Lazzati, D., et al., 2001, A\&A (in press), astro-ph/0109287

\bibitem[MacFayden \& Woosley 1999]{mw99}
MacFayden, A. \& Woosley, S. E., 1999, ApJ, 524, 262.

\bibitem[Madau, Pozzetti \& Dickinson 1998]{mpd98}
Madau, P., Pozzetti, L. \& Dickinson, M., 1998, ApJ, 498, 106.

\bibitem[Nakar \& Piran 2001]{np01}
Nakar, E. \&\ Piran, T. 2001, astro-ph/0103210.

\bibitem[Narayan, Piran \& Kumar 2001]{npk01}
Narayan, R., Piran, T. \& Kumar, P. 2001, ApJ 557, 949.

\bibitem[Oke \& Gunn 1983]{og83}
Oke, J.B. \& Gunn, J.E., 1983, ApJ, 266, 713.

\bibitem[Oke et al. 1995]{occ+95}
Oke, J.B., et al., 1995, PASP, 107, 375.

\bibitem[Panaitescu 2001]{panaitescu01}
Panaitescu, A., 2001, ApJ (accepted), astro-ph/0102401.

\bibitem[Panaitescu \& Kumar 2000]{pk01}
Panaitescu, A. \& Kumar, P., 2001, ApJ (submitted), astro-ph/0108045.

\bibitem[van Paradijs et~al. 1997]{paradijs+97}
van Paradijs J., et al., 1997, Nature, 386, 686.

\bibitem[Reichart, 1999]{reichart99}
Reichart, D.E., 1999, ApJ, 521, L111.

\bibitem[Sagar et al. 2001]{sagar+01}
Sagar, R., et al., 2001, BASI (submitted), astro-ph/0104249.

\bibitem[Sari, Piran \& Halpern 1999]{sph99}
Sari, R., Piran, T. \& Halpern, J., 1999, ApJ, 519, L17.

\bibitem[Sari, Piran \& Narayan 1998]{spn98}
Sari, R., Piran, T. \& Narayan, R., 1998, ApJ, 497, L17.

\bibitem[Schlegel, Finkbeiner \& Davis 1998]{sfd98}
Schlegel, D.J., Finkbeiner, D.P., \& Davis, M., 1998, ApJ, 500, 525.


\bibitem[van Putten 2001]{vanP01}
van Putten, M. H. P. M. 2001,  astro-ph/0109429 

\bibitem[Walker 1998]{walker98}
Walker, M.A., 1998, MNRAS, 294, 307.

\end{thebibliography}
\end{document}